# Osmosis through a Semi-permeable Membrane: a Consistent Approach to Interactions

Shixin Xu[1]
Bob Eisenberg[1,3,4]
Zilong Song[2]
Huaxiong Huang[1,2]

*June 7, 2018*

[1]The Fields Institute for Research in Mathematical Sciences, Toronto, ON, M5T 3J1, Canada
[2]Department of Mathematics & Statistics York University, Toronto, ON, M3J 1P3, Canada
[3]Department of Applied Mathematics, Illinois Institute of Technology, Chicago, IL, 60616, USA
[4]Department of Physiology & Biophysics, Rush University, Chicago, IL, 60612, USA

# Abstract

The movement of ionic solutions is an essential part of biology and technology. Fluidics, from nano- to micro- to microfluidics, is a burgeoning area of technology which is all about the movement of ionic solutions, on various scales. Many cells, tissues, and organs of animals and plants depend on osmosis, as the movement of fluids is called in biology. Indeed, the movement of fluids through channel proteins (that have a hole down their middle) is fluidics on an atomic scale. Ionic fluids are complex fluids, with energy stored in many ways. Ionic fluids flow driven by gradients of concentration, chemical and electrical potential, and hydrostatic pressure. Each flow is classically described by its own field theory, independent of the others, but of course, in reality every gradient drives every kind of flow to a varying extent. Combining field equations is tricky and so the theory of complex fluids derives the equations, rather than assumes their interactions. When field equations are derived, rather than assumed, their variables are consistent. That is to say all variables satisfy all equations under all conditions with one set of parameters. Here we treat a classical osmotic cell in this spirit, using a sharp interface method to derive interface boundary conditions consistent with all flows and fields. We allow volume to change with concentration, since changes of volume are a property of ionic solutions known to all who make them in the laboratory. We consider flexible and inflexible membranes. We show how to combine the energetics of the membrane with the energetics of the surrounding complex fluids. The results seem general but need application to specific situations of technological, biological and experimental importance before the consequences of consistency can be understood.

# Introduction

Osmosis moves ionic solutions throughout biology and technology. It is hard to find a more widespread phenomenon. All biology occurs in ionic solutions[1, 2] that move on many scales, including atomic scales smaller than nanometers, and a great deal of chemistry involves ionic movement as well. The modern technology of fluidics—macro, micro and nano—usually moves water and ions and thus involves osmosis at every scale. Indeed, modern names " ∗fluidics " may be viewed as a renaming of a classical,[3] if not ancient word[4] into more modern language.

Ionic solutions involve energy stored in many forms, pressure, concentration, electric and electrochemical potential, steric interactions, and chemical energy.[1] Ionic solutions are idealized as perfect gases[5] in elementary textbooks, whether biological,[6] technological (polarography)[7,8] or general.[9-16] Ionic solutions idealized this way do not have stored energy. The idealizations also do not involve time (in most cases) so they cannot deal with flow or friction.

Ionic solutions are complex fluids,[17-19] not simple at all.

Classical analysis of ionic solutions started with a thermodynamic and thus equilibrium analysis without flows.[11, 16] Thermodynamics was dynamic in its historical importance as a turning point in the history of science[20] but it was not dynamic—it did not involve time— in its physics or mathematics. Without time, it could not include flow or friction, at all. Flows occur in almost all living systems, and technological devices, so thermodynamics was extended *ad hoc* to deal with flows,[21-24] often with great success.[25-30]

The theory of complex fluids is a dynamic theory that always includes flows of many types, usually in a coherent consistent way. ('Consistent' means that all variables satisfy all equations and all boundary conditions with one set of parameters). In complex fluids like electrolytes, ions interact and move as components of complex fluids.[18, 31-37] The solution itself flows more or less as water itself would move (without the ions). Water molecules and ions move (partly) by bulk flow, that is to say, they move (partly) by convection described classically by Navier Stokes equations. Water molecules and ions also move (partly) by diffusion. Ions move (partly) because of the electric field. Water also moves in an electric field because of dielectrophoresis[38, 39] when the electric field is nonuniform i.e., $|\nabla E| > 0$, where $E$ is the electric field. Each of these flows varies with location and is described by field equations—typically partial differential equations in space and time—along with boundary conditions, that idealize the physics of the particular setup in which the flows occurs. One example of these field equations are the Maxwell equations of electrodynamics.[40-42] Few are more important[41, 43]. Most of our technology[44-49] is built from the Maxwell equations, perhaps because they are so precise and universal.

But even Maxwell's equations need to be extended to deal with osmosis. The Maxwell's equations do not involve hydrostatic pressure or concentrations of chemical species, for example, and ***so they give the same results, no matter what the pressure or concentration gradient.*** Thus, Maxwell's equations in themselves cannot deal with many crucial phenomena in biology and technology.

---

[1] We define 'chemical' forces between atoms as those that significantly change the spatial distribution of electrons in the atoms. In this definition, dielectric interactions are not a chemical force. More precisely, force produced by the induced charge, that is proportional to the local electric field, is not a chemical force. Dispersion forces arise from the quantum fluctuations in induced ('dielectric') forces and so are not considered chemical in this definition.

Pressure and concentration are responsible for many phenomena, in the laboratory, in technology and in biology.[21-24] Experiments show that gradients of hydrostatic pressure and concentration change flows a great deal, quantitatively and qualitatively in physical and living systems.[11, 21-24, 50] Maxwell's field equations cannot change with changes in hydrostatic pressure or flow because the field equations do not contain variables describing pressure or flow. Classical electrodynamics must be extended if it is to deal with the range of forces and flows so important in technology and life.

Conservation laws easily define the field equations of osmosis when a single force is involved, with some help from constitutive laws and boundary conditions (that describe the experimental or natural setup, more than anything else).[51, 52] But the situation is more complex when diffusion, and electrical migration, and bulk flow are all involved.[53] Bulk flow moves ions and water. Diffusion changes electrical potentials, electrical migration changes diffusion, and potential and diffusion (even of nonelectrolytes and water itself) change and are changed by bulk flow. Technology and biology use ionic solutions in devices where bulk flow, diffusion, and electrical migration are important. A theory must then deal with most of these conditions, robustly with one set of parameters, if it is to be useful. Only robust theories can create technologies as powerful as our semiconductor electronics. We reach to that success[54-56] as our goal.

Coupled field equations that satisfy all conservation laws with one set of parameters are consistent equations. Writing consistent equations is difficult. Even choosing the appropriate variables causes problems: for example, the operational definition of appropriate variables for an infinitely dilute sodium chloride solution remains requires extensive discussion[57] without definite resolution (after nearly a century of work).

Field equations have different form when different variables are chosen. Equations are typically written in terms of the mass of each specific chemical species, for example in terms of the mass of sodium and the mass of chloride in a salt solution. But a linear transformation would allow the equations to be written in terms of total mass density and net charge density, instead of sodium and chloride concentrations. The equations, coupling terms, and boundary conditions would be different. Choosing the 'right' set of equations is difficult.

Experimental results do not depend on how we define variables or formulate theories. The goal of theoretical analysis should be robust transferrable theories—that fit a range of data over a range of conditions with one set of parameters. The semiconductor technology which has remade our world is catalyzed by its successful theories. Similar robust theories of osmosis are likely to catalyze applications in biology and with substantial effects.

The theory of complex fluids has developed consistent theories for many complex systems[35, 58-60] that are mixtures (or even ionic solutions) with components that store energy. Variational methods deal successfully with magnetohydrodynamics systems[61], liquid crystals, polymeric fluids[36, 37], colloids and suspensions[58, 62] and electrorheological fluids[63]. Variational methods describe solid balls in liquids; deformable electrolyte droplets that fission and fuse[58, 64]; and suspensions of ellipsoids, including interfacial properties of these complex mixtures, such as surface tension and Marangoni effects of 'oil on water' and 'tears of wine'.[58, 62, 65, 66]

The Energetic Variational Approach (EnVarA) of Chun Liu, et. al., has been applied to ionic solutions of interest in references.[19, 67-72] EnVarA combines the Least Action Principle of Hamiltonian dynamics with the Maximum Dissipation Principle of Onsager (who used the dissipation function of Rayleigh) into a set of field equations, typically in Eulerian coordinates. Variations are taken with respect to two different variables, position and flow rate. The resulting field equations are written in one set of coordinates—usually Eulerian—with push-forward and pull-back methods.

EnVarA is an approach, not a general method. An approach adapts to the experimental setup and questions of interest. It avoids the abstraction of a universal formulation that may prove hard to understand[73-76] and hard to apply uniquely in specific cases.

The theory of complex fluids requires models that describe the constraints put on the field equations by biology, technology or experiments, as well as the field equations themselves. It requires boundary conditions as well as field equations. The constraints and boundary conditions often are the dominant determinant of the dynamics of the whole system. Describing these constraints with mathematical boundary conditions can be challenging, as in the moving contact-line problem[77-79] and when describing classical setups of the theory of stochastic processes.[80-84]

Boundary (or interface boundary) constraints of the physical problem can be converted into mathematical boundary conditions using the energy variational method.[78,79,85-87] The maximum dissipation principle (attributed to Onsager[88, 89 20-22] developed from the work of Rayleigh [74,75,90,91]) is often used to describe sharp interfaces.[18-19] Here we combine the energy variational approach (to field equations) with the sharp interface approach to physical constraints and use the combination to analyze a classical osmotic cell[3, 51, 92]—two baths separated by a semipermeable membrane—with flows driven by electrical, diffusion, and pressure fields, neglecting for the moment the steric and chemical forces of non-ideal solutions, or heat driven flows. Liu and colleagues have had some success creating what we call 'thermal dynamics' that deals consistently with heat driven flows in the spirit of EnVarA and the theory of complex fluids.[93-95]

The variational approach is widely used to analyze complex fluids because it derives field equations and boundary (or interface boundary) conditions, rather than assumes them. Once energy and dissipation functionals are defined by a physical model, the interaction terms in the field equations for each flow (bulk flow, diffusional flow, and electrical migration) are determined by algebra and analysis with minimal adjustable parameters. All variables satisfy all equations with one set of parameters when the analysis is done correctly. Results fit experiments, over a range of conditions with one set of parameters, if the original functional equations—from which the field equations are derived—are a correct enough model of the physics of the system, and the description of the constraints and boundary conditions is adequate. Results of the theory are then transferrable (using the language of chemistry[96,97]). Technology shows the importance of transferrable results. Our electronic/semiconductor technology would hardly exist if devices had different properties in each location in the remarkably diverse circuits they use.[44]

It is important to understand that field equations and boundary (or interface) conditions written without variational methods are likely to be incomplete, truncated, and thus not transferrable. Mathematical models written without variational methods are likely to omit terms needed to satisfy all field equations and boundary (or interface) constraints with one set of parameters, particularly if they are written as equations with right hand sides equal to zero, as is often the custom. (Vanishing terms are easy to overlook when applying conservation principles!) When the omitted terms have important effects, the truncated field equations cannot satisfy all field equations with one set of parameters. They are inconsistent. As the system is transferred from one set of conditions to another parameters must be changed to fit the data, and the new parameters are often only known by the data that is fit by them.

Our analysis of osmotic flow is in the spirit of a variational approach, combining elements of EnVarA and a sharp interface analysis that itself uses variational ideas. We also deal explicitly with the variation of the density of a solution with concentration,[98] responsible for the difference between molal and molar concentration units, taught to everyone who makes up solutions (from solid solutes) in a laboratory.[11, 99] This density dependence may be important in some experimental setups and conditions.

Our approach needs to be applied to each experimental setup before one knows its importance: different setups will emphasize different terms. Many theories are found to apply only in special setups and a particular range of conditions.[7,8,100-102] Disagreements are found in the literature of osmotic flow—some are mentioned in the Historical Note of this paper. Perhaps some of these disagreements reflect the use of different inconsistent theories—which may omit different terms—to deal with different experimental setups and conditions. Many setups will be sensitive to density changes with concentration: molar vs. molal terms may be important quite generally. See our eq. (23)-(24). We are unaware of other consistent analyses of the density effect—that is so well known to chemists and biologists who have made solutions in their laboratories—although the literature is vast and so is our lack of knowledge of it all.

A consistent theory should resolve differences, if the underlying physical model is correct.[2] (For us the underlying physical model is written in functional equations and boundary (or interface) constraints, e.g., eq. (1)-(2) that follow.) We hasten to add that none of us are physical chemists, let alone experimental physical chemists, and we are certainly aware that there are many experimental issues involved we do not know much about and that need to be included in applications of our work.

The paper is organized as follows. The mathematical model for a semi-permeable membrane is developed in Section 1. The development depends on several unknown variables. In Section 2, those variables are developed for the case where solution density is a function of ion concentration. Those unknown variables are evaluated for the case of constant density in Section 3. A historical section and discussion conclude the paper.

**<u>Note to the reader</u>.** Variational approaches have not been as widely used as they might be, perhaps because they involve a large amount of tricky mathematics that has to be done and checked. We present all steps in our analysis, hoping to make the approach easier to use. Apologies are extended to those who find the detail overwhelming, or irritating.

---

[2] For us the underlying physical model is written in functional equations and boundary (or interface) constraints, e.g., eq. (1)-(2) below.

# Mathematical Model

We consider the dynamics of a fluid with ions passing through a seim-permeable membrane in the traditional setup of physical chemists and biologists[21-24, 50, 92, 103] following [104, 105]. We use the sharp interface model[77, 78, 85, 86, 106] to derive detailed specific conditions on the membrane. Let $\Omega^\pm$ denote two compartments separated by a membrane $\Gamma$ where $\Omega = \Omega^+ \cup \Omega^-$. (see Fig. 1). $\mathbf{D}^\pm$ is the electric displacement vector field of classical electrodynamics and $c_i^\pm$ is the distribution of the of $i_{th}$ ($i = 1, \ldots, N$) species of ion in domain $\Omega^\pm$, respectively. $\rho^\pm$ and $\mathbf{u}^\pm$ are the density and velocity of solutions in left and right compartments, respectively.

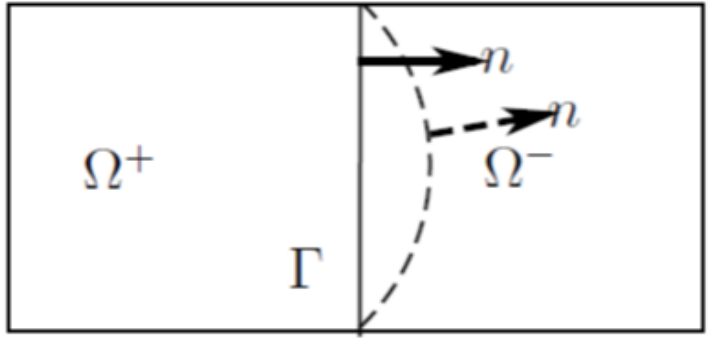

**Figure 1**: Schematic of a flexible membrane. Solid line is initial shape of membrane and the dashed line denotes the deformed membrane. $\boldsymbol{n}$ is the normal vector of membrane $\Gamma$ from the left compartment $\Omega^+$ to the right compartment $\Omega^-$.

Based on the Maxwell's equations and conservation law for the mass of each ionic species, we have the following equations

$$\begin{cases} \nabla \cdot \mathbf{D}^\pm = \sum_i z_i e c_i^\pm, & in\ \Omega^\pm \\ \dfrac{\partial c_i^\pm}{\partial t} + \nabla \cdot (c_i^\pm \boldsymbol{u}^\pm) = -\nabla \cdot \mathbf{j}_i^\pm, & in\ \Omega^\pm \\ \rho^\pm \left( \dfrac{\partial \boldsymbol{u}^\pm}{\partial t} + (\boldsymbol{u}^\pm \cdot \nabla) \boldsymbol{u}^\pm \right) = \nabla \cdot \left( \boldsymbol{\sigma}_\eta^\pm + \boldsymbol{\sigma}_e^\pm \right), & in\ \Omega^\pm \\ \dfrac{\partial \rho^\pm}{\partial t} + \nabla \cdot (\rho^\pm \boldsymbol{u}^\pm) = 0, & in\ \Omega^\pm \end{cases} \quad (1)$$

$z_i$ is the valence of the $i_{th}$ species, $e$ is the elementary charge, $\mathbf{j}_i^\pm$ is the flux of the $i_{th}$ species $\boldsymbol{\sigma}_\eta^\pm$ is the viscous stress, and $\boldsymbol{\sigma}_e^\pm$ is the electric stress in domain $\Omega^\pm$, respectively. $\boldsymbol{\sigma}_\eta^\pm$ and $\boldsymbol{\sigma}_e^\pm$ are symetric tensors.

If we neglect magnetic forces, there exists an electric field $\mathbf{E}^\pm$ and electrical potential $\phi^\pm$ such that the dielectric constant $\mathbf{D}^\pm = \varepsilon_0 \varepsilon_r^\pm \mathbf{E}^\pm = -\varepsilon_0 \varepsilon_r^\pm \nabla \phi^\pm$, where $\varepsilon_0 \varepsilon_r^\pm$ is in domain $\Omega^\pm$, respectively. We assume that the dielectric constant $\varepsilon_r^\pm$ (units: dimensionless) and the permittivity $\varepsilon_0 \varepsilon_r^\pm$ (units: farads/meter) are each a single real number.

We do not deal with the nonideal properties of ionic solutions, despite our understanding of their significance, and the importance of nonideality in general.[1, 43, 107, 108] We have to start somewhere and we have our hands full with the mathematics needed to deal consistently with the idealized cases. When we do reach to include nonideality, we anticipate difficulties. Even the proper formulation of the field equations (with flow) in the nonideal case remains an open question. If we neglect the magnetic forces, there exist an electric field $\boldsymbol{E}^{\pm}$ and electric potential $\phi^{\pm}$ such that $\boldsymbol{D}^{\pm} = \varepsilon_0 \varepsilon_r^{\pm} \boldsymbol{E}^{\pm} = -\varepsilon_0 \varepsilon_r^{\pm} \nabla \phi^{\pm}$, where $\varepsilon_0 \varepsilon_r^{\pm}$ is the dielectric constant in domain $\Omega^{\pm}$, respectively. We assume the dielectric constant $\varepsilon_r^{\pm}$ (units: dimensionless) and the permittivity $\varepsilon_0 \varepsilon_r^{\pm}$ (units: farads/meter) are each a single real number and do not deal with the nonideal properties of ionic solutions, despite our understanding of their significance, and the importance of nonideality in general[1,43,107,108] We have to start somewhere and we have our hands full with the mathematics needed to deal consistently with these idealized cases. When we do reach to include nonideality, we anticipate difficulties. Even the proper formulation of the field equations (with flow) in the nonideal case remains an open question.

For the interface condition of the surface $\Gamma$, also based on conservation law[109], we assume that

$$\text{All on } \Gamma \begin{cases} \dfrac{d\mathbf{X}}{dt} = \boldsymbol{v}, \\ \boldsymbol{D}^{\pm} \cdot \boldsymbol{n} = Q(x), \\ \rho^{\pm} \boldsymbol{u}^{\pm} \cdot \boldsymbol{n} - \rho^{\pm} \boldsymbol{v} \cdot \boldsymbol{n} = Q_{\rho}, \\ \boldsymbol{u}^{\pm} \cdot \boldsymbol{\tau} = \boldsymbol{v} \cdot \boldsymbol{\tau}, \\ [\boldsymbol{\sigma}_{\eta} + \boldsymbol{\sigma}_e] \cdot \boldsymbol{n} - [Q_{\rho} \boldsymbol{u}] = \boldsymbol{F}_{mb} \\ \mathbf{j}_i^{\pm} \cdot \boldsymbol{n} + c_i^{\pm}(\boldsymbol{u}^{\pm} - \boldsymbol{v}) \cdot \boldsymbol{n} = J_i(x) \end{cases} \quad (2)$$

where $\mathbf{X}(\cdot, t)$ is the trajectories of points on the membrane, $\boldsymbol{v} = \boldsymbol{v}_{\tau} + v_n \boldsymbol{n}$ is the velocity of membrane with normal vector $\boldsymbol{n}$ and tangential vector $\boldsymbol{\tau}$, $[f] = f^+ - f^-$ is the difference of $f$ across the interface $\Gamma$. We also assume that $\mathbf{u}^{\pm} = \boldsymbol{v} = 0$ on $\partial \Gamma$, i.e., we assume the edge (or boundary) of membrane is fixed. Here $Q(x)$, $Q_{\rho}(x)$, and $J_i(x)$ are surface charge density, solution flux and ion flux on the interface, respectively. $\boldsymbol{F}_{mb}$ is the membrane force induced by mechanical and dielectric properties of membrane.

In the next section, we will derive the explicit forms of $\mathbf{j}_i^{\pm}$, $\boldsymbol{\sigma}_e^{\pm}$, $Q(x)$, $Q_{\rho}$, $J_i(x)$ and $\boldsymbol{F}_{mb}$ based on the second law of thermodynamics generalized to deal with flows driven by different forces.

**Remark** *$\boldsymbol{F}_{mb}$ is taken as a surface force for simplicity. However, it can include the other types of forces such as elastic force and/or dissipation force induced by friction inside of membrane.*

**Total Energy Functional.** In our case, and many others,[58, 78, 85, 86, 106] the total energy functional is the sum of the kinetic energy part $E_{kin}(\rho, \boldsymbol{u})$, internal energy $E_{int}(c_i, \phi)$ and surface energy $E_\Gamma(\phi, \gamma_0)$. The internal energy is composed of the electrostatic energy part $E_{es}(\phi)$, and the Gibbs free energy of ions $E_{ion}(c_i)$. The total energy functional is the precise statement **of our field model** (that is to say our physical model without boundary ($\partial\Omega$) constraints or conditions). And the total energy functional is given by

$$E^{tot} = E_{kin} + E_{int} + E_\Gamma$$
$$= E_{kin} + E_{es} + E_{ion} + E_\Gamma$$
$$= \sum_\pm \int_{\Omega^\pm} (e_{kin}^\pm + e_{es}^\pm + e_{ion}^\pm) dx + \int_\Gamma (e_\Gamma + \gamma_0) dS \qquad (3)$$
$$= \sum_\pm \int_{\Omega^\pm} \left\{ \frac{1}{2}\rho^\pm |\boldsymbol{u}^\pm|^2 + \frac{1}{2}\mathbf{E}^\pm \cdot \mathbf{D}^\pm + k_B T \sum_i c_i^\pm \ln\left(\frac{c_i^\pm}{c_0}\right) \right\} dx$$
$$+ \int_\Gamma \left( \frac{C_m}{2}[\phi]^2 + \gamma_0 \right) dS$$

where $c_0$ is a characteristic ion density, $C_m$ is membrane capacitance, $\gamma_0$ is the membrane surface tension. In the following, square brackets always denote the jumps across the interface. $\sum_\pm \int_{\Omega^\pm} f^\pm dx = \int_{\Omega^+} f^+ dx + \int_{\Omega^-} f^- dx$ is the sum of integration in left and right compartments weighted by the corresponding value of $f$.

The **dissipation functional** is defined as

$$\Delta = \sum_\pm \int_{\Omega^\pm} 2\eta^\pm |\boldsymbol{D}_\eta^\pm|^2 dx + \sum_\pm \int_{\Omega^\pm} \lambda^\pm |\nabla \cdot \boldsymbol{u}^\pm|^2 dx + \frac{1}{k_B T} \sum_\pm \int_{\Omega^\pm} \sum_i D_i^\pm c_i^\pm |\nabla \tilde{\mu}_i^\pm|^2 dx$$
$$+ \int_\Gamma G_1([\tilde{\mu}_i]) dS + \int_\Gamma G_2(Q_\rho) dS \qquad (4)$$

Where $\boldsymbol{D}_\eta = (\nabla \boldsymbol{u} + (\nabla \boldsymbol{u})^T)/2$ is rate of strain, $\eta^\pm$ and $\lambda^\pm$ are the two Lame constants[109], $\tilde{\mu}_i^\pm = \tilde{\mu}_i^\pm(c_i^\pm, \phi^\pm, \rho^\pm)$ and $D_i^\pm$ are the chemical potential and diffusion coefficient of $i_{th}$ ion. The first three terms in dissipation functional are the dissipation induced by fluid friction, volume change and ion diffusion in the bulk region. The last two terms are the dissipation induced by irreversible osmosis .on the membrane. The forms of $G_1(x) \geq 0$ and $G_2(x) \geq 0$, for any $x$, will be discussed later.

# Derivation: Density Depends on Concentration

We start with fluid density as a function of ion density $\rho = \hat{\rho}(c_1^\pm, c_2^\pm, \cdots c_N^\pm)$. Based on the biological applications, here we consider the case that $\rho$ weakly depends on the the ion concentration.

## Permeable and Flexible Membranes

In this section, we consider the membrane is deformable (see Fig. 1). We first present a generalized Reynolds transfer formula[110, 111] when the membrane is permeable

$$\frac{d}{dt} \sum_\pm \int_{\Omega^\pm(t)} f dx = \sum_\pm \int_{\Omega^\pm(t)} \frac{\partial f^\pm}{\partial t} dx + \int_\Gamma [f] \boldsymbol{v} \cdot \boldsymbol{n} dS \quad (5)$$

$$= \sum_\pm \int_{\Omega^\pm(t)} \left( \frac{df^\pm}{dt} + f^\pm \nabla \cdot \boldsymbol{u}^\pm \right) dx - \int_\Gamma \left[\frac{f}{\rho}\right] Q_\rho dS - \int_{\partial\Omega} f^\pm \boldsymbol{u}^\pm \cdot \boldsymbol{n} \, dS,$$

where $\int_{\partial\Omega(t)} f^\pm dS = \int_{\partial\Omega \cap \bar{\Omega}^+} f^+ dS + \int_{\partial\Omega \cap \bar{\Omega}^-} f^- dS$.

Based on the results in,[77, 112, 113] we have the following formula for surface defined function $f$,

$$\frac{d}{dt} \int_{\Gamma(t)} f dS = \int_{\Gamma(t)} (\dot{f} + f \nabla_\Gamma \cdot \boldsymbol{v}) dS = \int_{\Gamma(t)} \left( \frac{\partial f}{\partial t} + \boldsymbol{v} \cdot \nabla f + f \nabla_\Gamma \cdot \boldsymbol{v} \right) d$$

$$= \int_{\Gamma(t)} \left( \frac{\partial f}{\partial t} + v_n \partial_n f + \boldsymbol{v}_\tau \cdot \nabla_\Gamma f + f \nabla_\Gamma \cdot \boldsymbol{v} \right)$$

$$= \int_{\Gamma(t)} \frac{\partial f}{\partial t} + v_n \partial_n f dS + \int_{\partial \Gamma(t)} f \boldsymbol{v}_\tau \cdot \boldsymbol{n}_m dl - \int_{\Gamma(t)} f H \boldsymbol{n} \cdot \boldsymbol{v} dS$$

$$= \int_{\Gamma(t)} \left(\frac{d^n f}{dt}\right) dS - \int_{\Gamma(t)} f H \boldsymbol{n} \cdot \boldsymbol{v} dS \quad (6)$$

where $\nabla_\Gamma = \nabla - \boldsymbol{n}(\boldsymbol{n} \cdot \nabla)$ is the surface gradient operator, $\dot{f} = \frac{\partial f}{\partial t} + \boldsymbol{v} \cdot \nabla f$ is the material derivative on the membrane, $\frac{d^n f}{dt} = \frac{\partial f}{\partial t} + v_n \partial_n f$ is the normal time derivative, $\boldsymbol{n}_m$ is the unit outward normal vector of membrane at the edge $\partial \Gamma$, i.e. $\boldsymbol{n}_m \cdot \boldsymbol{n} = 0$, $\boldsymbol{n}_m \cdot d\boldsymbol{l} = 0$ and $H = -\nabla \cdot \boldsymbol{n}$ is the mean curve of membrane $\Gamma$. Here we used the fact that $\boldsymbol{v} = \boldsymbol{0}$ on $\partial \Gamma$ in our model.

The following useful lemma states that the normal time derivative of normal vector on the surface only depends on normal component of membrane velocity.

**Lemma** Let $\mathbf{n}$ and $\mathbf{v} = v_n \mathbf{n} + \mathbf{v}_\tau$ be the outward normal vector and velocity of membrane $\Gamma$, respectively. Then we have the following result

$$\frac{d^n \mathbf{n}}{dt} = -\nabla_\Gamma v_n.$$

The proof of **Lemma** is presented in Appendix A.

Now we begin to derive the full model. If we take the derivative of the energy (3),

$$\frac{dE^{tot}}{dt} = \frac{dE_{kin}}{dt} + \frac{dE_{es}}{dt} + \frac{dE_{ion}}{dt} + \frac{dE_\Gamma}{dt} = I_1 + I_2 + I_3 + I_4 \tag{7}$$

For the first term $I_1$, by using the last two equations in Eq. (1), third interface condtion in (2) and Eq. (5), we have

$$I_1 = \frac{d}{dt} \sum_\pm \int_{\Omega^\pm} \frac{1}{2}\rho^\pm |\mathbf{u}^\pm|^2 dx$$

$$= \sum_\pm \int_{\Omega^\pm} \frac{|\mathbf{u}^\pm|^2}{2} \frac{d\rho^\pm}{dt} dx + \sum_\pm \int_{\Omega^\pm} \rho^\pm \frac{d\mathbf{u}^\pm}{dt} \cdot \mathbf{u}^\pm dx - \int_\Gamma \left[\frac{|\mathbf{u}|^2}{2}\right] Q_\rho$$

$$+ \sum_\pm \int_{\Omega^\pm} \frac{1}{2}\rho^\pm |\mathbf{u}^\pm|^2 \nabla \cdot \mathbf{u}^\pm dx - \int_{\partial\Omega} e^\pm_{kin} \mathbf{u}^\pm \cdot \mathbf{n} dS$$

$$= -\sum_\pm \int_{\Omega^\pm} \boldsymbol{\sigma}^\pm_\eta : \nabla \mathbf{u}^\pm dx - \sum_\pm \int_{\Omega^\pm} \boldsymbol{\sigma}^\pm_e : \nabla \mathbf{u}^\pm dx + \int_\Gamma [\mathbf{n} \cdot (\boldsymbol{\sigma}_\eta + \boldsymbol{\sigma}_e) \cdot \mathbf{u}] dS$$

$$- \int_\Gamma \left[\frac{|\mathbf{u}|^2}{2}\right] Q_\rho dS + I_{1b}.$$

$$= -\sum_\pm \int_{\Omega^\pm} \boldsymbol{\sigma}^\pm_\eta : \nabla \mathbf{u}^\pm dx - \sum_\pm \int_{\Omega^\pm} \boldsymbol{\sigma}^\pm_e : \nabla \mathbf{u}^\pm dx + \int_\Gamma [(\mathbf{n} \cdot (\boldsymbol{\sigma}_\eta + \boldsymbol{\sigma}_e) - Q_\rho \mathbf{u}) \cdot \mathbf{u}] dS$$

$$\tag{8}$$

$$+ \int_\Gamma \left[\frac{u_n^2}{2}\right] Q_\rho dS + \sum_\pm \int_{\Omega^\pm} p^\pm_c \left(\sum_i \frac{\partial_i \rho^\pm}{\rho^\pm} (\nabla \cdot \mathbf{j}^\pm_i + c^\pm_i \nabla \cdot \mathbf{u}^\pm) - \nabla \cdot \mathbf{u}^\pm\right) dx + I_{1b.}$$

where we used the notation $\partial_i \rho^\pm = \frac{\partial \hat{\rho}}{\partial c_i}(c^\pm_i)$, the decompostion $\mathbf{u} = u_n \mathbf{n} + \mathbf{u}_\tau$, $\mathbf{u}_\tau$ is continuous on the membrane and

$$I_{1b} = \int_{\partial\Omega} \left((\boldsymbol{\sigma}^\pm_\eta + \boldsymbol{\sigma}^\pm_e) \cdot \mathbf{n}\right) \cdot \mathbf{u}^\pm dS - \int_{\partial\Omega} e^\pm_{kin} \mathbf{u}^\pm \cdot \mathbf{n} dS.$$

The colon **:** represents the double dot of two tensors $\mathbf{A}:\mathbf{B} \equiv \sum_{ij} A_{ij} B_{ij}$.

A Lagrange multiplier $p_c$ is introduced to assure mass conservation, when solution density is a function of ion density.

The second and third terms in Eq. (7) are , with the help of first two equations in (1) and last two equations in (2), given by

$$I_2 + I_3 = \frac{d}{dt}\sum_{\pm}\int_{\Omega^{\pm}} e_{es}^{\pm}\,dx + \frac{d}{dt}\sum_{\pm}\int_{\Omega^{\pm}} e_{ion}^{\pm}\,dx$$

$$= \sum_{\pm}\int_{\Omega^{\pm}} \frac{\partial}{\partial t}\left(\frac{1}{2}\,\boldsymbol{E}^{\pm}\cdot\boldsymbol{D}^{\pm}\right)dx + \sum_{\pm}\int_{\Omega^{\pm}} \frac{\partial}{\partial t}\left(\sum_i k_B T c_i^{\pm}\ln\frac{c_i^{\pm}}{c_0}\right)dx + \int_{\Gamma}[e_{es}+e_{ion}]\boldsymbol{v}\cdot\boldsymbol{n}\,dS$$

$$= \sum_{\pm}\int_{\Omega^{\pm}}\sum_i\left(z_i e\phi^{\pm} + k_B T\left(\ln\left(\frac{c_i^{\pm}}{c_0}\right)+1\right)\right)\frac{\partial c_i^{\pm}}{\partial t}\,dx$$

$$-\int_{\partial\Omega}\left(\phi^{\pm}\frac{\partial \boldsymbol{D}^{\pm}}{\partial t}\cdot\boldsymbol{n}\right)dS - \int_{\Gamma}\left[\phi\frac{\partial\boldsymbol{D}}{\partial t}\cdot\boldsymbol{n}\right]dS + \int_{\Gamma}[e_{es}+e_{ion}]\boldsymbol{v}\cdot\boldsymbol{n}\,dS$$

$$= \sum_{\pm}\int_{\Omega^{\pm}}\sum_i \mu_i^{\pm}\left(-\nabla\cdot\boldsymbol{j}_i^{\pm} - \nabla\cdot(c_i^{\pm}\boldsymbol{u}^{\pm})\right)dx$$

$$-\int_{\partial\Omega}\left(\phi^{\pm}\frac{\partial\boldsymbol{D}^{\pm}}{\partial t}\cdot\boldsymbol{n}\right)dS - \int_{\Gamma}\left[\phi\frac{\partial\boldsymbol{D}}{\partial t}\cdot\boldsymbol{n}\right]dS + \int_{\Gamma}[e_{es}+e_{ion}]\boldsymbol{v}\cdot\boldsymbol{n}\,dS$$

$$= \sum_{\pm}\int_{\Omega^{\pm}}\sum_i \nabla\mu_i^{\pm}\cdot\boldsymbol{j}_i^{\pm}\,dx + \sum_{\pm}\int_{\Omega^{\pm}}\sum_i c_i^{\pm}(\nabla\mu_i^{\pm})\cdot\boldsymbol{u}^{\pm}\,dx$$

$$-\int_{\Gamma}\left[\sum_i \mu_i(\boldsymbol{j}_i\cdot\boldsymbol{n} + c_i\boldsymbol{u}\cdot\boldsymbol{n})\right]dS$$

$$-\int_{\Gamma}\left[\phi\frac{\partial\boldsymbol{D}}{\partial t}\cdot\boldsymbol{n}\right]dS + \int_{\Gamma}[e_{es}+e_{ion}]\boldsymbol{v}\cdot\boldsymbol{n}\,dS + I_{21b}$$

$$= \sum_{\pm}\int_{\Omega^{\pm}}\sum_i \nabla\mu_i^{\pm}\cdot\boldsymbol{j}_i^{\pm}\,dx + \sum_{\pm}\int_{\Omega^{\pm}}\sum_i c_i^{\pm}(\nabla\mu_i^{\pm})\cdot\boldsymbol{u}^{\pm}\,dx \qquad (9)$$

$$-\int_{\Gamma}\sum_i [\mu_i]J_i\,dS - \int_{\Gamma}\left[\phi\frac{\partial\boldsymbol{D}}{\partial t}\cdot\boldsymbol{n}\right]dS + \int_{\Gamma}\left[\left(e_{es}+e_{ion} - \sum_i \mu_i c_i\right)\right]\boldsymbol{v}\cdot\boldsymbol{n}\,dS + I_{21b}$$

where $\mu_i^{\pm} = z_i e\phi^{\pm} + k_B T\left(\ln\left(\frac{c_i^{\pm}}{c_0}\right)+1\right)$ and

$$I_{21b} = -\int_{\partial\Omega}\left(\phi^{\pm}\frac{\partial \boldsymbol{D}^{\pm}}{\partial t}\cdot\boldsymbol{n}\right)dS - \int_{\partial\Omega}\mu_i\left(j_i^{\pm}+c_i^{\pm}\boldsymbol{u}^{\pm}\right)\cdot\boldsymbol{n}\,dS.$$

Due to the first equation in (1) and second equation in (2), the second term in Eq. (9) is rewritten

$$\sum_{\pm}\int_{\Omega^{\pm}}\sum_{i}c_i^{\pm}\nabla\mu_i^{\pm}\cdot\boldsymbol{u}^{\pm}dx =$$

$$=\sum_{\pm}\int_{\Omega^{\pm}}\sum_{i}z_i e c_i^{\pm}\nabla\phi^{\pm}\cdot\boldsymbol{u}^{\pm}dx + \sum_{\pm}\int_{\Omega^{\pm}}\sum_{i}k_B T \nabla c_i^{\pm}\cdot\boldsymbol{u}^{\pm}\,dx$$

$$=\sum_{\pm}\int_{\Omega^{\pm}}(\nabla\cdot\boldsymbol{D}^{\pm})(\nabla\phi^{\pm}\cdot\boldsymbol{u}^{\pm})dx - \sum_{\pm}\int_{\Omega^{\pm}}\sum_{i}k_B T c_i^{\pm}\nabla\cdot\boldsymbol{u}^{\pm}\,dx$$

$$+\int_{\partial\Omega}\sum_{i}k_B T c_i^{\pm}\boldsymbol{u}^{\pm}\cdot\boldsymbol{n}\,dS + \int_{\Gamma}\left[\sum_{i}k_B T c_i\,\boldsymbol{u}\cdot\boldsymbol{n}\right]dS$$

$$=-\sum_{\pm}\int_{\Omega^{\pm}}\boldsymbol{D}^{\pm}\cdot(\nabla\boldsymbol{u}^{\pm})\cdot\nabla\phi^{\pm}dx - \sum_{\pm}\int_{\Omega^{\pm}}\boldsymbol{D}^{\pm}\cdot(\nabla\nabla\phi^{\pm})\cdot\boldsymbol{u}^{\pm}dx + \int_{\Gamma}[Q\nabla\phi\cdot\boldsymbol{u}]\,dS$$

$$+\int_{\partial\Omega}(\boldsymbol{D}^{\pm}\cdot\boldsymbol{n})(\nabla\phi^{\pm}\cdot\boldsymbol{u}^{\pm})dS$$

$$-\sum_{\pm}\int_{\Omega^{\pm}}\sum_{i}k_B T c_i^{\pm}\nabla\cdot\boldsymbol{u}^{\pm}dx + \int_{\partial\Omega}\sum_{i}k_B T c_i^{\pm}\boldsymbol{u}^{\pm}\cdot\boldsymbol{n}\,dS + \int_{\Gamma}\left[\sum_{i}k_B T c_i\,\boldsymbol{u}\cdot\boldsymbol{n}\right]dS$$

$$=\sum_{\pm}\int_{\Omega^{\pm}}\epsilon_0\epsilon_r^{\pm}(\nabla\phi^{\pm}\otimes\nabla\phi^{\pm}):\nabla\boldsymbol{u}^{\pm}dx + \sum_{\pm}\int_{\Omega^{\pm}}\frac{\epsilon_0\epsilon_r^{\pm}}{2}(\nabla|\nabla\phi^{\pm}|^2)\cdot\boldsymbol{u}^{\pm}\,dx$$

$$+\int_{\Gamma}[Q\nabla\phi\cdot\boldsymbol{u}]dS + \int_{\partial\Omega}(\boldsymbol{D}^{\pm}\cdot\boldsymbol{n})(\nabla\phi^{\pm}\cdot\boldsymbol{u}^{\pm})\,dS$$

$$-\sum_{\pm}\int_{\Omega^{\pm}}\sum_{i}k_B T c_i^{\pm}\nabla\cdot\boldsymbol{u}^{\pm}dx + \int_{\partial\Omega}\sum_{i}k_B T c_i^{\pm}\boldsymbol{u}^{\pm}\cdot\boldsymbol{n}\,dS + \int_{\Gamma}\left[\sum_{i}k_B T c_i\,\boldsymbol{u}\cdot\boldsymbol{n}\right]dS$$

$$=\sum_{\pm}\int_{\Omega^{\pm}}\epsilon_0\epsilon_r^{\pm}\left(\nabla\phi^{\pm}\otimes\nabla\phi^{\pm}-\frac{|\nabla\phi^{\pm}|^2}{2}\boldsymbol{I}\right):\nabla\boldsymbol{u}^{\pm}dx - \sum_{\pm}\int_{\Omega^{\pm}}\sum_{i}k_B T c_i^{\pm}\nabla\cdot\boldsymbol{u}^{\pm}\,dx$$

$$+\int_{\Gamma}[e_{es}\boldsymbol{u}\cdot\boldsymbol{n}]dS + \int_{\Gamma}[Q\nabla\phi\cdot\boldsymbol{u}]dS + \int_{\Gamma}\left[\sum_{i}k_B T c_i\boldsymbol{u}\cdot\boldsymbol{n}\right]dS + I_{22b},$$

where

$$I_{22b} = \int_{\partial\Omega} e_{es}^{\pm} \boldsymbol{u}^{\pm} \cdot \boldsymbol{n} dS + \int_{\partial\Omega} (\boldsymbol{D}^{\pm} \cdot \boldsymbol{n})(\nabla\phi^{\pm} \cdot \boldsymbol{u}^{\pm}) dS + \int_{\partial\Omega} \sum_i k_B T c_i^{\pm} \boldsymbol{u} \cdot \boldsymbol{n} \, dS. \qquad (10)$$

By using the **Lemma** on the interface $\Gamma(t)$, we have

$$\frac{d^n Q}{dt} = \frac{d^n \boldsymbol{D}}{dt} \cdot \boldsymbol{n} + \frac{d^n \boldsymbol{n}}{dt} \cdot \boldsymbol{D} = \frac{\partial \boldsymbol{D}}{\partial t} \cdot \boldsymbol{n} + v_n \partial_n \boldsymbol{D} \cdot \boldsymbol{n} - \nabla_\Gamma v_n \cdot \boldsymbol{D},$$

which yields

$$\frac{\partial \boldsymbol{D}}{\partial t} \cdot \boldsymbol{n} = \frac{d^n Q}{dt} + \nabla_\Gamma v_n \cdot \boldsymbol{D} - v_n \partial_n \boldsymbol{D} \cdot \boldsymbol{n} = \frac{d^n Q}{dt} - \epsilon_0 \epsilon_r \nabla_\Gamma v_n \cdot \nabla_\Gamma \phi - v_n \partial_n \boldsymbol{D} \cdot \boldsymbol{n}.$$

Then the fourth term in (9) could be rewritten as

$$\int_\Gamma \left[\phi \frac{\partial \boldsymbol{D}}{\partial t} \cdot \boldsymbol{n}\right] dS = \int_\Gamma [\phi] \frac{d^n Q}{dt} dS - \int_\Gamma [\epsilon_0 \epsilon_r \phi \nabla_\Gamma v_n \cdot \nabla_\Gamma \phi + \phi v_n \partial_n \boldsymbol{D} \cdot \boldsymbol{n}] dS \qquad (11)$$

$$= \int_\Gamma [\phi] \frac{d^n Q}{dt} dS + \int_\Gamma [\epsilon_0 \epsilon_r \nabla_\Gamma \cdot (\phi \nabla_\Gamma \phi)] v_n \, dS - \int_\Gamma [\phi \partial_n \boldsymbol{D} \cdot \boldsymbol{n}] v_n \, dS.$$

Combining (9), (10) and (11) yields

$$I_2 + I_3 =$$

$$= \sum_\pm \int_{\Omega^\pm} \sum_i \nabla \mu_i^\pm \cdot \boldsymbol{j}_i^\pm dx + \sum_\pm \int_{\Omega^\pm} \epsilon_0 \epsilon_r^\pm \left(\nabla \phi^\pm \otimes \nabla \phi^\pm - \frac{|\nabla \phi^\pm|^2}{2} \boldsymbol{I}\right) : \nabla \boldsymbol{u}^\pm dx$$

$$- \sum_\pm \int_{\Omega^\pm} \sum_i k_B T c_i^\pm \nabla \cdot \boldsymbol{u}^\pm dx + \int_\Gamma \left[\left(e_{es} + \sum_i k_B T c_i\right) \boldsymbol{u} \cdot \boldsymbol{n}\right] dS + \int_\Gamma [Q \nabla \phi \cdot \boldsymbol{u}] dS$$

$$- \int_\Gamma \sum_i [\mu_i] J_i \, dS - \int_\Gamma [\phi] \frac{d^n Q}{dt} dS \qquad (12)$$

$$- \int_\Gamma \left[\nabla_\Gamma \cdot \left(\nabla_\Gamma \left(\frac{\epsilon_0 \epsilon_r}{2} \phi^2\right)\right)\right] \boldsymbol{v} \cdot \boldsymbol{n} dS + \int_\Gamma [\phi \partial_n \boldsymbol{D} \cdot \boldsymbol{n}] \boldsymbol{v} \cdot \boldsymbol{n} \, dS$$

$$+ \int_\Gamma [e_{es} + e_{ion} - c_i \mu_i] \boldsymbol{v} \cdot \boldsymbol{n} dS + I_{2b}$$

where $I_{2b} = I_{21b} + I_{22b}$.

For the last term in Eq. (7), by using the surface Reynolds formula (6), we obtain

$$I_4 = \int_\Gamma C_m [\phi] \frac{d^n}{dt}([\phi]) \, dS - \int_\Gamma ((e_\Gamma + \gamma_0) H \boldsymbol{n}) \cdot \boldsymbol{v} dS. \qquad (13)$$

Combining Eqs.(8), (12) and (13), we have

$$\frac{dE^{tot}}{dt} =$$

$$= -\sum_i \int_{\Omega^\pm} \boldsymbol{\sigma}_\eta^\pm : \nabla \boldsymbol{u}^\pm dx$$

$$- \sum_\pm \int_{\Omega^\pm} \left( \boldsymbol{\sigma}_e^\pm - \epsilon_0 \epsilon_r^\pm \left( \nabla \phi^\pm \otimes \nabla \phi^\pm - \frac{|\nabla \phi^\pm|^2}{2} \boldsymbol{I} \right) \right) : \nabla \boldsymbol{u}^\pm dx$$

$$+ \sum_\pm \int_{\Omega^\pm} \sum_i \nabla \mu_i^\pm \cdot \boldsymbol{j}_i^\pm dx - \sum_\pm \int_{\Omega^\pm} \sum_i k_B T c_i^\pm \nabla \cdot \boldsymbol{u}^\pm dx$$

$$+ \sum_\pm \int_{\Omega^\pm} p_c^\pm \left( \sum_i \frac{\partial_i \rho^\pm}{\rho^\pm} (\nabla \cdot \boldsymbol{j}_i^\pm + c_i^\pm \nabla \cdot \boldsymbol{u}^\pm) - \nabla \cdot \boldsymbol{u}^\pm \right) dx$$

$$+ \int_\Gamma [(\boldsymbol{n} \cdot (\boldsymbol{\sigma}_\eta + \boldsymbol{\sigma}_e) - Q_\rho \boldsymbol{u}) \cdot \boldsymbol{u}] dS + \int_\Gamma \left[ \left( e_{es} + \sum_i k_B T c_i \right) \boldsymbol{u} \cdot \boldsymbol{n} \right] dS$$

$$+ \int_\Gamma [Q \nabla \phi \cdot \boldsymbol{u}] dS - \int_\Gamma \sum_i [\mu_i] J_i \, dS - \int_\Gamma [\phi] \frac{d^n Q}{dt} dS$$

$$- \int_\Gamma \left[ \nabla_\Gamma \cdot \left( \nabla_\Gamma \left( \frac{\epsilon_0 \epsilon_r}{2} \phi^2 \right) \right) \right] \boldsymbol{v} \cdot \boldsymbol{n} dS + \int_\Gamma [\phi \partial_n \boldsymbol{D} \cdot \boldsymbol{n}] \boldsymbol{v} \cdot \boldsymbol{n} \, dS$$

$$+ \int_\Gamma \left[ \left( e_{es} + e_{ion} - \sum_i \mu_i c_i \right) \right] \boldsymbol{v} \cdot \boldsymbol{n} dS + \int_\Gamma \left[ \frac{|u_n|^2}{2} \right] Q_\rho dS$$

$$+ \int_\Gamma C_m[\phi] \frac{d^n}{dt} ([\phi]) \, dS - \int_\Gamma (e_\Gamma + \gamma_0) H \boldsymbol{v} \cdot \boldsymbol{n} \, dS + I_{1b} + I_{2b} \tag{14}$$

By using the fact that

$$\sum_\pm \int_{\Omega^\pm} \left( p_c^\pm \sum_i \frac{\partial_i \rho^\pm}{\rho^\pm} \nabla \cdot \boldsymbol{j}_i^\pm \right) dx =$$

$$= -\sum_\pm \int_{\Omega^\pm} \nabla \left( p_c^\pm \sum_i \frac{\partial_i \rho^\pm}{\rho^\pm} \right) \cdot \boldsymbol{j}_i^\pm dx + \int_\Gamma \left[ \left( p_c \sum_i \frac{\partial_i \rho}{\rho} \right) \boldsymbol{j}_i \cdot \boldsymbol{n} \right] dS$$

$$+ \int_{\partial \Omega} \left( p_c^\pm \sum_i \frac{\partial_i \rho^\pm}{\rho^\pm} \right) \boldsymbol{j}_i^\pm \cdot \boldsymbol{n} \, dS$$

$$= -\sum_\pm \int_{\Omega^\pm} \nabla \left( p_c^\pm \sum_i \frac{\partial_i \rho^\pm}{\rho^\pm} \right) \cdot \boldsymbol{j}_i^\pm dx + \int_\Gamma \left[ p_c \sum_i \frac{\partial_i \rho}{\rho} \left( J_i - \frac{c_i}{\rho} Q_\rho \right) \right] dS$$

$$+ \int_{\partial\Omega} \left( p_c^{\pm} \sum_i \frac{\partial_i \rho^{\pm}}{\rho^{\pm}} \right) \boldsymbol{j}_i^{\pm} \cdot \boldsymbol{n} \, dS$$

$$= -\sum_{\pm} \int_{\Omega^{\pm}} \nabla \left( p_c^{\pm} \sum_i \frac{\partial_i \rho^{\pm}}{\rho^{\pm}} \right) \cdot \boldsymbol{j}_i^{\pm} dx + \int_{\Gamma} \left[ \sum_i p_c \frac{\partial_i \rho}{\rho} \right] J_i \, dS \qquad (15)$$

$$- \int_{\Gamma} \left[ \sum_i p_c \frac{\partial_i \rho}{\rho} \frac{c_i}{\rho} \right] Q_\rho \, dS + \int_{\partial\Omega} \left( p_c^{\pm} \sum_i \frac{\partial_i \rho^{\pm}}{\rho^{\pm}} \right) \boldsymbol{j}_i^{\pm} \cdot \boldsymbol{n} \, dS.$$

Eq. (14) can be written as

$$\frac{dE^{tot}}{dt} = -\sum_i \int_{\Omega^{\pm}} \boldsymbol{\sigma}_{\eta}^{\pm} : \nabla \boldsymbol{u}^{\pm} dx$$

$$- \sum_{\pm} \int_{\Omega^{\pm}} \left( \boldsymbol{\sigma}_e^{\pm} - \epsilon_0 \epsilon_r^{\pm} \left( \nabla \phi^{\pm} \otimes \nabla \phi^{\pm} - \frac{|\nabla \phi^{\pm}|^2}{2} \boldsymbol{I} \right) \right) : \nabla \boldsymbol{u}^{\pm} \, dx$$

$$+ \sum_{\pm} \int_{\Omega^{\pm}} \sum_i \nabla \tilde{\mu}_i^{\pm} \cdot \boldsymbol{j}_i^{\pm} dx - \int_{\Gamma} \sum_i [\tilde{\mu}_i] J_i \, dS$$

$$+ \sum_{\pm} \int_{\Omega^{\pm}} \left( -\sum_i k_B T c_i^{\pm} - p_c^{\pm} \left( 1 - \sum_i \frac{\partial_i \rho^{\pm}}{\rho^{\pm}} c_i \right) \right) \nabla \cdot \boldsymbol{u}^{\pm} \, dx$$

$$+ \int_{\Gamma} \left[ \left( (\boldsymbol{\sigma}_\eta + \boldsymbol{\sigma}_e) \cdot \boldsymbol{n} - Q_\rho \boldsymbol{u} + \left( e_{es} + \sum_i k_B T c_i \right) \boldsymbol{n} + Q \nabla \phi \right) \cdot \boldsymbol{u} \right] dS$$

$$+ \int_{\Gamma} \left[ e_{es} + e_{ion} - \sum_i \mu_i c_i + \phi \partial_n \boldsymbol{D} \cdot \boldsymbol{n} - \nabla_{\Gamma} \cdot \left( \nabla_{\Gamma} \left( \frac{\epsilon_0 \epsilon_r}{2} \phi^2 \right) \right) \right] \boldsymbol{v} \cdot \boldsymbol{n} \, dS$$

$$+ \int_{\Gamma} \left[ -p_c \sum_i \frac{\partial_i \rho}{\rho^2} c_i + \frac{|u_n|^2}{2} \right] Q_\rho dS$$

$$+ \int_{\Gamma} [\phi] \frac{d^n}{dt} (C_m [\phi] - Q) \, dS - \int_{\Gamma} (e_{\Gamma} + \gamma_0) H \boldsymbol{v} \cdot \boldsymbol{n} \, dS + I_b$$

$$= -\sum_i \int_{\Omega^{\pm}} \boldsymbol{\sigma}_\eta^{\pm} : \nabla \boldsymbol{u}^{\pm} dx - \sum_{\pm} \int_{\Omega^{\pm}} \left( \boldsymbol{\sigma}_e^{\pm} - \epsilon_0 \epsilon_r^{\pm} \left( \nabla \phi^{\pm} \otimes \nabla \phi^{\pm} - \frac{|\nabla \phi^{\pm}|^2}{2} \boldsymbol{I} \right) \right) : \nabla \boldsymbol{u}^{\pm} \, dx$$

$$+ \sum_{\pm} \int_{\Omega^{\pm}} \sum_i \nabla \tilde{\mu}_i^{\pm} \cdot \boldsymbol{j}_i^{\pm} dx - \int_{\Gamma} \sum_i [\tilde{\mu}_i] J_i \, dS$$

$$+ \sum_{\pm} \int_{\Omega^{\pm}} \left( -\sum_i k_B T c_i^{\pm} - p_c^{\pm} \left( 1 - \sum_i \frac{\partial_i \rho^{\pm}}{\rho^{\pm}} c_i \right) \right) \nabla \cdot \boldsymbol{u}^{\pm} \, dx$$

$$+ \int_{\Gamma} \left\{ \left[ (\boldsymbol{\sigma}_\eta + \boldsymbol{\sigma}_e) \cdot \boldsymbol{n} - Q_\rho \boldsymbol{u} + \left( e_{es} + \sum_i k_B T c_i \right) \boldsymbol{n} + Q \nabla \phi \right] \right\} \cdot \boldsymbol{v}_\tau \, dS$$

$$+ \int_{\Gamma} \left[ \left( (\boldsymbol{\sigma}_\eta + \boldsymbol{\sigma}_e) \cdot \boldsymbol{n} - Q_\rho \boldsymbol{u} + \left( e_{es} + \sum_i k_B T c_i \right) \boldsymbol{n} + Q \nabla \phi \right) \cdot \boldsymbol{n} u_n \right] dS$$

$$+ \int_{\Gamma} \left[ e_{es} + e_{ion} - \sum_i \mu_i c_i + \phi \partial_n \boldsymbol{D} \cdot \boldsymbol{n} - \nabla_\Gamma \cdot \left( \nabla_\Gamma \left( \frac{\epsilon_0 \epsilon_r}{2} \phi^2 \right) \right) \right] \boldsymbol{v} \cdot \boldsymbol{n} \, dS$$

$$+ \int_{\Gamma} \left[ -p_c \sum_i \frac{\partial_i \rho}{\rho^2} c_i + \frac{|u_n|^2}{2} \right] Q_\rho dS$$

$$+ \int_{\Gamma} [\phi] \frac{d^n}{dt} (C_m[\phi] - Q) \, dS - \int_{\Gamma} (e_\Gamma + \gamma_0) H \boldsymbol{v} \cdot \boldsymbol{n} \, dS + I_b$$

$$= - \sum_i \int_{\Omega^\pm} \boldsymbol{\sigma}_\eta^\pm : \nabla \boldsymbol{u}^\pm \, dx$$

$$- \sum_\pm \int_{\Omega^\pm} \left( \boldsymbol{\sigma}_e^\pm - \epsilon_0 \epsilon_r^\pm \left( \nabla \phi^\pm \otimes \nabla \phi^\pm - \frac{|\nabla \phi^\pm|^2}{2} \boldsymbol{I} \right) \right) : \nabla \boldsymbol{u}^\pm \, dx$$

$$+ \sum_\pm \int_{\Omega^\pm} \sum_i \nabla \tilde{\mu}_i^\pm \cdot \boldsymbol{j}_i^\pm dx \; - \int_{\Gamma} \sum_i [\tilde{\mu}_i] J_i \, dS$$

$$+ \sum_\pm \int_{\Omega^\pm} \left( - \sum_i k_B T c_i^\pm - p_c^\pm \left( 1 - \sum_i \frac{\partial_i \rho^\pm}{\rho^\pm} c_i \right) \right) \nabla \cdot \boldsymbol{u}^\pm dx$$

$$+ \int_{\Gamma} (\boldsymbol{F}_{mb} + [Q \nabla \phi]) \cdot \boldsymbol{v}_\tau dS$$

$$+ \int_{\Gamma} \left[ \frac{((\boldsymbol{\sigma}_\eta + \boldsymbol{\sigma}_e) \cdot \boldsymbol{n} - Q_\rho \boldsymbol{u} + (e_{es} + \sum_i k_B T c_i) \boldsymbol{n} + Q \nabla \phi)}{\rho} \right] \cdot \boldsymbol{n} Q_\rho \, dS$$

$$+ \int_{\Gamma} \left( \boldsymbol{F}_{mb} + \left[ \left( e_{es} + \sum_i k_B T c_i \right) \boldsymbol{n} + Q \nabla \phi \right] \right) \cdot \boldsymbol{n} v_n \, dS$$

$$+ \int_{\Gamma} \left[ e_{es} + e_{ion} - \sum_i \mu_i c_i + \phi \partial_n \boldsymbol{D} \cdot \boldsymbol{n} - \nabla_\Gamma \cdot \left( \nabla_\Gamma \left( \frac{\epsilon_0 \epsilon_r}{2} \phi^2 \right) \right) \right] v_n \, dS$$

$$+ \int_{\Gamma} \left[ -p_c \sum_i \frac{\partial_i \rho}{\rho^2} c_i + \frac{|u_n|^2}{2} \right] Q_\rho \, dS$$

$$+ \int_{\Gamma} [\phi] \frac{d^n}{dt} (C_m[\phi] - Q) \, dS - \int_{\Gamma} (e_\Gamma + \gamma_0) H v_n \, dS + I_b$$

$$
\begin{aligned}
= & -\sum_i \int_{\Omega^\pm} \boldsymbol{\sigma}_\eta^\pm : \nabla \boldsymbol{u}^\pm \, dx \\
& -\sum_\pm \int_{\Omega^\pm} \left( \boldsymbol{\sigma}_e^\pm - \epsilon_0 \epsilon_r^\pm \left( \nabla \phi^\pm \otimes \nabla \phi^\pm - \frac{|\nabla \phi^\pm|^2}{2} \boldsymbol{I} \right) \right) : \nabla \boldsymbol{u}^\pm \, dx \\
& + \sum_\pm \int_{\Omega^\pm} \sum_i \nabla \tilde{\mu}_i^\pm \cdot \boldsymbol{j}_i^\pm dx \; - \int_\Gamma \sum_i [\tilde{\mu}_i] J_i \, dS \\
& + \sum_\pm \int_{\Omega^\pm} \left( -\sum_i k_B T c_i^\pm - p_c^\pm \left( 1 - \sum_i \frac{\partial_i \rho^\pm}{\rho^\pm} c_i \right) \right) \nabla \cdot \boldsymbol{u}^\pm \, dx \\
& + \int_\Gamma (\boldsymbol{F}_{mb} + [Q \nabla \phi]) \cdot \boldsymbol{v}_\tau dS \\
& + \int_\Gamma (\boldsymbol{F}_{mb} \cdot \boldsymbol{n} + [F_n] - (e_\Gamma + \gamma_0) H) v_n dS \\
& + \int_\Gamma \left[\frac{F_\rho}{\rho}\right] Q_\rho dS + \int_\Gamma [\phi] \frac{d^n}{dt}(C_m[\phi] - Q) \, dS - I_b
\end{aligned}
\quad (16)
$$

where $\tilde{\mu}_i^\pm = \mu_i^\pm - p_c^\pm \frac{\partial_i \rho^\pm}{\rho^\pm}$, $I_b = I_{1b} + I_{2b} + \int_{\partial \Omega} \left( p_c^\pm \sum_i \frac{\partial_i \rho^\pm}{\rho^\pm} \right) \boldsymbol{j}_i^\pm \cdot \boldsymbol{n} \, dS$

$$
\begin{aligned}
F_n &= Q \nabla \phi \cdot \boldsymbol{n} + \left( \sum_i k_B T c_i + 2 e_{es} + e_{ion} - \sum_i c_i \mu_i \right) - \nabla_\Gamma \cdot \left( \nabla_\Gamma \left( \frac{\epsilon_0 \epsilon_r}{2} \phi^2 \right) \right) + \phi (\partial_n \boldsymbol{D} \cdot \boldsymbol{n}) \\
&= Q \, \partial_n \phi + \left( 2 e_{es} - \sum_i c_i z_i e \phi \right) - \nabla_\Gamma \cdot \left( \nabla_\Gamma \left( \frac{\epsilon_0 \epsilon_r}{2} \phi^2 \right) \right) + \phi (\partial_n \boldsymbol{D} \cdot \boldsymbol{n}) \\
&= Q \partial_n \phi + \left( \epsilon_0 \epsilon_r |\nabla_\Gamma \phi|^2 + \epsilon_0 \epsilon_r |\partial_n \phi|^2 - \sum_i c_i z_i e \phi \right) \\
& \quad - (\epsilon_0 \epsilon_r |\nabla_\Gamma \phi|^2 + \epsilon_0 \epsilon_r \phi \Delta_\Gamma \phi) + \phi (\partial_n \boldsymbol{D} \cdot \boldsymbol{n}) \\
&= -\left( \sum_i c_i z_i e \phi + \epsilon_0 \epsilon_r \phi \Delta_\Gamma \phi - \phi \partial_n \boldsymbol{D} \cdot \boldsymbol{n} \right) \\
&= -(\phi \nabla \cdot \boldsymbol{D} + \epsilon_0 \epsilon_r \phi \Delta_\Gamma \phi - \phi \partial_n \boldsymbol{D} \cdot \boldsymbol{n}) \\
&= -(\phi \partial_n \boldsymbol{D} \cdot \boldsymbol{n} + \phi \nabla_\Gamma \cdot \boldsymbol{D} + \epsilon_0 \epsilon_r \phi \Delta_\Gamma \phi - \phi \partial_n \boldsymbol{D} \cdot \boldsymbol{n}) \\
&= -(\phi \nabla_\Gamma \cdot \boldsymbol{D}_\Gamma - \boldsymbol{D} \cdot \boldsymbol{n} \, \phi H + \epsilon_0 \epsilon_r \phi \Delta_\Gamma \phi) \\
&= Q \phi H
\end{aligned}
\quad (17)
$$

and

$$
F_\rho = -Q_\rho u_n + \boldsymbol{n} \cdot (\boldsymbol{\sigma}_\eta + \boldsymbol{\sigma}_e) \cdot \boldsymbol{n} + \left( e_{es} + \sum_i k_B T c_i \right) + Q \partial_n \phi - \sum_i p_c \frac{\partial_i \rho}{\rho} c_i + \frac{\rho |u_n|^2}{2} \quad (18)
$$

Comparing with Eq. (4), by taking quadratic forms of $G_1 = \sum_i \frac{g_i}{e^2}[\tilde{\mu}_i]^2$ and $G_2 = \frac{Q_\rho^2}{K(x)}$, we have

$$\begin{cases} \boldsymbol{j}_i^\pm = -\frac{D_i^\pm c_i^\pm}{K_B T}\nabla\tilde{\mu}_i^\pm & \text{in } \Omega^\pm \\ & \text{in } \Omega^\pm \\ \boldsymbol{\sigma}_\eta^\pm = 2\eta^\pm \boldsymbol{D}_\eta^\pm + \lambda^\pm(\nabla\cdot\boldsymbol{u}^\pm)\boldsymbol{I} - p\boldsymbol{I} & \text{in } \Omega^\pm i_i \\ \boldsymbol{\sigma}_e^\pm = \varepsilon_0\varepsilon_r^\pm \nabla\phi^\pm \otimes \nabla\phi^\pm - \frac{\varepsilon_0\varepsilon_r^\pm}{2}|\nabla\phi^\pm|^2\boldsymbol{I} & \text{in } \Omega^\pm \\ J_i = \frac{g_i}{(z_i e)^2}[\tilde{\mu}_i] & \text{in } \Omega^\pm \\ p^\pm = \sum_i k_B T c_i^\pm + p_c^\pm\left(1 - \sum_i \frac{\partial_i \rho^\pm}{\rho^\pm}c_i^\pm\right) & \text{on } \Gamma \\ Q = C_m[\phi] & \text{on } \Gamma \\ Q_\rho = -\bar{\rho}^2 K(\boldsymbol{x})\left\{\left[\frac{F_\rho}{\rho}\right]\right\} & \text{on } \Gamma \\ \boldsymbol{F}_{mb}\cdot\boldsymbol{n} = (e_\Gamma + \gamma_0)H - [F_n] & \text{on } \Gamma \\ \boldsymbol{\tau}\cdot\boldsymbol{F}_{mb} = -Q[\nabla_\Gamma\phi]\cdot\boldsymbol{\tau} & \end{cases} \quad (19)$$

where $K(x)$ and $g_i$ are the permeability and conductance of the membrane, respectively.

**Remark** *Here we took $G_1$ and $G_2$ as the simple qudratic functions, i.e. $G_1 = \sum_i(g_i/e^2)[\tilde{\mu}_i]^2$ and $G_2 = Q_\rho^2/K(x)$.* **In fact, any flux $J(x)$ satisfying $G(x) = xJ(x) \geq 0$ could be used in order to** *maximize the dissipation.*

Then, we have

$$[F_n] = Q[\phi]H = 2e_\Gamma H \text{ and } -Q[\nabla_\Gamma\phi] = -\nabla_\Gamma e_\Gamma \quad (20)$$

which yields

$$[(\boldsymbol{\sigma}_\eta + \boldsymbol{\sigma}_e)]\cdot\boldsymbol{n} - [Q_\rho\boldsymbol{u}] = \boldsymbol{F}_{mb} = (\gamma_0 - e_\Gamma)H\boldsymbol{n} - \nabla_\Gamma e_\Gamma. \quad (21)$$

and

$$\left[\frac{F_\rho}{\rho}\right] = \left[\frac{1}{\rho}\left(\boldsymbol{n}\cdot(\boldsymbol{\sigma}_\eta + \boldsymbol{\sigma}_e)\cdot\boldsymbol{n} + \left(e_{es} + \sum_i k_B T c_i\right) + Q\partial_n\phi - \sum_i p_c\frac{\partial_i\rho}{\rho}c_i\right)\right]$$

$$+ \left[\frac{|u_n|^2}{2} - \frac{Q_\rho u_n}{\rho}\right]$$

$$= \left[\frac{1}{\rho}\left(\boldsymbol{n}\cdot(\boldsymbol{\sigma}_\eta)\cdot\boldsymbol{n} + 0 - \sum_i p_c\frac{\partial_i\rho}{\rho}c_i\right)\right] + \left[\frac{Q_\rho^2 + 2Q_\rho v_n}{2\rho^2} - \frac{Q_\rho}{\rho}\left(\frac{Q_\rho}{\rho} + v_n\right)\right]$$

$$= \left[\frac{1}{\rho}\left(\boldsymbol{n}\cdot(\boldsymbol{\sigma}_\eta)\cdot\boldsymbol{n} + 0 - \sum_i p_c\frac{\partial_i\rho}{\rho}c_i\right)\right] - \left[\frac{Q_\rho^2}{2\rho^2}\right]. \tag{22}$$

Here we used the fact that $v_n$ is continous across the membrane.

**Model for fluid with variable density.** To summarize, we have the following model for fluid with variable density passing through a deformable membrane

$$\begin{cases} -\nabla\cdot(\varepsilon_0\varepsilon_r^\pm\nabla\phi^\pm) = \sum_i z_i e c_i^\pm, & \text{in } \Omega^\pm \\[4pt] \dfrac{\partial c_i^\pm}{\partial t} + \nabla\cdot(\boldsymbol{u}^\pm c_i^\pm) = \nabla\cdot\left(D_i^\pm\left(\nabla c_i^\pm + \dfrac{z_i e}{k_B T}c_i^\pm\nabla\phi^\pm - \dfrac{1}{k_B T}c_i^\pm\nabla\left(p_c^\pm\dfrac{\partial_i\rho^\pm}{\rho^\pm}\right)\right)\right), & \text{in } \Omega^\pm \\[4pt] \rho^\pm\dfrac{d\boldsymbol{u}^\pm}{dt} = \nabla\cdot(\boldsymbol{\sigma}_\eta^\pm) + \nabla\cdot(\boldsymbol{\sigma}_e^\pm), & \text{in } \Omega^\pm \\[4pt] \dfrac{d\rho^\pm}{dt} + \rho^\pm\nabla\cdot\boldsymbol{u}^\pm = 0, & \text{in } \Omega^\pm \\[4pt] \rho = \hat{\rho}(c_1^\pm, c_2^\pm, \cdots c_N^\pm) & \text{in } \Omega^\pm \end{cases} \tag{23}$$

where viscosity shear stress is

$$\boldsymbol{\sigma}_\eta^\pm = 2\eta^\pm \boldsymbol{D}_\eta^\pm + \lambda^\pm(\nabla\cdot\boldsymbol{u}^\pm)\boldsymbol{I} - \left(\sum_i k_B T c_i^\pm + p_c^\pm\left(1 - \sum_i \frac{\partial_i\rho^\pm}{\rho^\pm}c_i\right)\right)\boldsymbol{I}$$

and Maxwell stress $\boldsymbol{\sigma}_e^\pm = \epsilon_0\epsilon_r^\pm(\nabla\phi^\pm\otimes\nabla\phi^\pm - \frac{1}{2}|\nabla\phi|^2\boldsymbol{I})$. with interface conditions on $\Gamma$

$$\begin{cases} \dfrac{d\boldsymbol{X}}{dt} = \boldsymbol{v} \\[4pt] -\varepsilon_0\varepsilon_r^\pm\nabla\phi^\pm\cdot\boldsymbol{n} = C_m[\phi], \\[4pt] \boldsymbol{j}_i^\pm\cdot\boldsymbol{n} + c_i^\pm(\boldsymbol{u}^\pm - \boldsymbol{v})\cdot\boldsymbol{n} = \dfrac{g_i}{(z_i e)^2}\left(\left[k_B T(\ln(c_i/c_0) + 1) + z_i e\phi - p_c\dfrac{\partial_i\rho}{\rho}\right]\right), \\[4pt] Q_\rho = -\bar{\rho}^2 K(\boldsymbol{x})\left\{\left[\dfrac{1}{\rho}\left(\boldsymbol{n}\cdot\boldsymbol{\sigma}_\eta\cdot\boldsymbol{n} + 0 - \sum_i p_c\dfrac{\partial_i\rho}{\rho}c_i\right)\right] - \left[\dfrac{1}{2\rho^2}\right]Q_\rho^2\right\} \\[4pt] \rho^\pm(\boldsymbol{u}^\pm - \boldsymbol{v})\cdot\boldsymbol{n} = Q_\rho \\[4pt] [\boldsymbol{\sigma}_\eta + \boldsymbol{\sigma}_e]\cdot\boldsymbol{n} - [Q_\rho\boldsymbol{u}] = (\gamma_0 - e_\Gamma)H\boldsymbol{n} - \nabla_\Gamma e_\Gamma \\[4pt] \boldsymbol{u}^\pm\cdot\boldsymbol{\tau} = \boldsymbol{v}\cdot\boldsymbol{\tau} \end{cases} \tag{24}$$

and some appropriate boundary conditions on $\partial\Omega$.

**Remark.** *Here it is easy to check that the model (23) with interface conditions (24) is Galilean invariant. For the transmembrane mass flux, note that (23) is a quadratic equation of $Q_\rho$, which normally involves two solutions. If we denote $\delta = [1/2\rho^2] \ll 1$, one solution involves $Q_\rho^2$ as a small perturbation*

$$Q_\rho = -\bar{\rho}^2 K(x) \left\{ \left[ \frac{1}{\rho} \left( \mathbf{n} \cdot \boldsymbol{\sigma}_\eta \cdot \mathbf{n} + O - \sum_i p_c \frac{\partial_i \rho}{\rho} c_i \right) \right] \right\}$$

$$+ \delta \bar{\rho}^2 K(x) \left\{ \bar{\rho}^2 K(x) \left[ \frac{1}{\rho} \left( \mathbf{n} \cdot \boldsymbol{\sigma}_\eta \cdot \mathbf{n} + O - \sum_i p_c \frac{\partial_i \rho}{\rho} c_i \right) \right] \right\}^2 + O(\delta^2),$$

*which means $Q_\rho$ is maily driven by the hydro and osmotic pressure difference. The other candidate solution makes $Q_\rho$ a large quantity*

$$Q_\rho = \frac{1}{\delta} \frac{1}{\bar{\rho}^2 K(x)} - \bar{\rho}^2 K(x) \left\{ \left[ \frac{1}{\rho} \left( \mathbf{n} \cdot \boldsymbol{\sigma}_\eta \cdot \mathbf{n} + O - \sum_i p_c \frac{\partial_i \rho}{\rho} c_i \right) \right] \right\} + O(\delta),$$

*which means $Q_\rho$ is mainly driven by the density difference and does not depend on the hydrostatic and osmotic pressure. The large $Q_\rho$ solution is not reasonable for a semi-permeable membrane flux because the mass flux increases as density difference decreases ($\delta \to 0$). So the first solution will be chosen for our semi-permeable membrane mass flux.*

**Remark** *Here we considered a simple membrane mechanical energy, only including the surface tension. The derivation could be generalized to **more complicated membrane energy case, for example the Helfrich bending energy**[87, 114] and the neo-Hookean hyperelastic energy[115], allowing models of fascinating biological phenomena like vesicle fusion[116-118] to include the effects of ionic composition, membrane potential, and hydrostatic pressure that are known to have significant effects in experiment and life. **Details are in the Appendix B.***

# Permeable and Inflexible Membranes

In this section, we assume that the membrane is inflexible $v = 0$ and the total energy and dissipation functional are same as in Eqs. (3) and (4). The flux interface conditions are reduced to $\rho^{\pm}\mathbf{u}^{\pm} = Q_{\rho}\mathbf{n}$ and $\mathbf{j}_i^{\pm} \cdot \mathbf{n} + c_i^{\pm}\mathbf{u}^{\pm} \cdot \mathbf{n} = J_i(x), on\ \Gamma$. By using the similar calculation with the reduced interface conditions, Eq. (16) is reduced to

$$\begin{aligned}
\frac{dE^{tot}}{dt} = &-\sum_{\pm}\int_{\Omega^{\pm}} \boldsymbol{\sigma}_{\eta}^{\pm} : \nabla \mathbf{u}^{\pm}\, dx \\
&- \sum_{\pm}\int_{\Omega^{\pm}} \left(\boldsymbol{\sigma}_e - \varepsilon_0 \varepsilon_r^{\pm}(\nabla\phi^{\pm} \otimes \nabla\phi^{\pm} - \frac{1}{2}|\nabla\phi^{\pm}|^2 \mathbf{I})\right) : \nabla \mathbf{u}\, dx \\
&+ \sum_{\pm}\int_{\Omega^{\pm}} \sum_i (\nabla\tilde{\mu}_i^{\pm}) \cdot \mathbf{j}_i^{\pm}\, dx - \int_{\Gamma}\sum_i [\tilde{\mu}_i] J_i\, dS \\
&+ \sum_{\pm}\int_{\Omega^{\pm}} \left(-\sum_i k_B T c_i^{\pm} - p_c^{\pm}\left(1 - \sum_i \frac{\partial_i \rho^{\pm}}{\rho^{\pm}} c_i\right)\right) \nabla \cdot \mathbf{u}^{\pm}\, dx \\
&+ \int_{\Gamma} [\phi] \frac{\partial}{\partial t}(C_m[\phi] - Q)\, dS + \int_{\Gamma} \left[\frac{F_{\rho}}{\rho}\right] Q_{\rho}\, dS + I_b.
\end{aligned} \quad (25)$$

Comparing with the dissipation functional (4), we obtain the same results as in Eq. (19) without the last two equations for force balance on the membrane.

**Model for an inflexible semi-permeable membrane.** To summarize, when the density of solution is a function of ion concentration, we obtained the following model for fluid passing through an inflexible semi-permeable membrane

$$\begin{cases}
-\nabla \cdot (\varepsilon_0 \varepsilon_r^{\pm} \nabla \phi^{\pm}) = \sum_i z_i e c_i^{\pm}, & in\ \Omega^{\pm} \\
\frac{\partial c_i^{\pm}}{\partial t} + \nabla \cdot (\mathbf{u}^{\pm} c_i^{\pm}) = \nabla \cdot \left(D_i^{\pm}\left(\nabla c_i^{\pm} + \frac{z_i e}{k_B T} c_i^{\pm} \nabla \phi^{\pm} - \frac{1}{k_B T} c_i^{\pm} \nabla \left(p_c^{\pm} \frac{\partial_i \rho^{\pm}}{\rho^{\pm}}\right)\right)\right), & in\ \Omega^{\pm} \\
\rho^{\pm} \frac{d\mathbf{u}^{\pm}}{dt} = \nabla \cdot (\boldsymbol{\sigma}_{\eta}^{\pm}) + \nabla \cdot \left(\varepsilon_0 \varepsilon_r^{\pm} \nabla \phi^{\pm} \otimes \nabla \phi^{\pm} - \frac{\varepsilon_0 \varepsilon_r^{\pm}}{2}|\nabla\phi^{\pm}|^2\right), & in\ \Omega^{\pm} \\
\frac{d\rho^{\pm}}{dt} + \rho^{\pm} \nabla \cdot \mathbf{u}^{\pm} = 0, & in\ \Omega^{\pm} \\
\rho = \hat{\rho}(c_1^{\pm}, c_2^{\pm}, \cdots c_N^{\pm}) & in\ \Omega^{\pm}
\end{cases} \quad (26)$$

with interface conditions on $\Gamma$

$$\begin{cases} -\varepsilon_0 \varepsilon_r^\pm \nabla \phi^\pm \cdot \mathbf{n} = C_m[\phi], \\[6pt] \mathbf{j}_i^\pm \cdot \mathbf{n} + c_i^\pm \mathbf{u}^\pm \cdot \mathbf{n} = \dfrac{g_i}{(z_i e)^2}\left(\left[k_B T(\ln(c_i/c_0)+1) + z_i e\phi - p_c \dfrac{\partial_i \rho}{\rho}\right]\right), \\[6pt] Q_\rho = -\bar{\rho}^2 K(\mathbf{x})\left\{\left[\dfrac{1}{\rho}\left(0 + \mathbf{n}\cdot\boldsymbol{\sigma}_\eta \cdot \mathbf{n} - \sum_i \dfrac{\partial_i \rho}{\rho} p_c c_i\right)\right] - \left[\dfrac{1}{2\rho^2}\right] Q_\rho^2\right\} \\[6pt] \rho^\pm \mathbf{u}^\pm = Q_\rho \mathbf{n}, \end{cases} \quad (27)$$

and some appropriate boundary conditions on $\partial\Omega$.

**Remark** *Comparing with the results for flexible membrane in Eqs. (23)-(24), we find the only difference is that there is no force balance equation for the inflexible membrane. The inflexible membrane has an unknown external force that maintains the position of membrane. Only the Dirichlet interface condition is used to describe the solvent velocity passing through the inflexible membrane. In the case that the membrane is non-permeable for fluid, an equation of osmotic velocity is not needed. In this case, the membrane moves with the fluid, i.e., $Q_\rho = 0$ (i.e. $\mathbf{u}^\pm = \mathbf{v}$); only the force balance equation $[\boldsymbol{\sigma}_\eta + \boldsymbol{\sigma}_e] \cdot \mathbf{n} = (\gamma_0 - e_\Gamma)H\mathbf{n} - \nabla_\Gamma e_\Gamma$ on the membrane is needed when the membrane is non-permeable for fluid.*

# Model with Constant Solution Density

For the normal biological problem, where the bulk ion concentration is in the range of $1\sim 300 mM$, the density of solution has small variation. Concentrations are very much larger in and near ion channels, charged lipid membranes, binding proteins, enzyme active sites, and nucleic acids. In those situations solutions are also quite nonideal and so a different treatment is needed.

In the bulk, we can take the zero order approximation, i.e. $\rho^\pm = \rho^0$, where $\rho^0$ is a constant. Then the fourth equation in (1) is reduced to the imcompressiblility condition $\nabla \cdot \mathbf{u} = 0$. The kinetic energy functional is reduced to be $E_{kin} = \sum_\pm \int_{\Omega^\pm} \frac{1}{2}\rho^0 |\mathbf{u}^\pm|^2 dx$. The second term in dissipation functional vanishes. The interface conditions are reduced to

$$\text{on } \Gamma \begin{cases} \dfrac{d\mathbf{X}}{dt} = \mathbf{v}, \\[6pt] \mathbf{D}^\pm \cdot \mathbf{n} = Q(x), \\[6pt] \mathbf{u}^\pm \cdot \mathbf{n} - \mathbf{v}\cdot \mathbf{n} = U, \\[6pt] \mathbf{j}_i^\pm \cdot \mathbf{n} + c_i^\pm(\mathbf{u}^\pm - \mathbf{v})\cdot \mathbf{n} = J_i(x), \\[6pt] [\boldsymbol{\sigma}_\eta + \boldsymbol{\sigma}_e]\cdot \mathbf{n} = \mathbf{F}_{mb} \\[6pt] \mathbf{u}^\pm \cdot \boldsymbol{\tau} = \mathbf{v}\cdot \boldsymbol{\tau} \end{cases} \quad (28)$$

Here the main difference between Eq. (2) and Eq. (29) is that the velocity on the membrane is continuous since the density is continuous. Here we replaced $\frac{Q_\rho}{\rho^0}$ by new variable $U$.

# Permeable and Flexible Membranes

We also first assume that the membrane is flexible. Then by using a similar calculation with interface conditions (28) and incompressibility, Eq. (16) is reduced to

$$\begin{aligned}
\frac{dE^{tot}}{dt} = &- \sum_i \int_{\Omega^\pm} \boldsymbol{\sigma}_\eta^\pm : \nabla \boldsymbol{u}^\pm dx \\
&- \sum_\pm \int_{\Omega^\pm} \left( \boldsymbol{\sigma}_e^\pm - \epsilon_0 \epsilon_r^\pm \left( \nabla \phi^\pm \otimes \nabla \phi^\pm - \frac{|\nabla \phi^\pm|^2}{2} \boldsymbol{I} \right) \right) : \nabla \boldsymbol{u}^\pm dx \\
&+ \sum_\pm \int_{\Omega^\pm} \sum_i \nabla \mu_i^\pm \cdot \boldsymbol{j}_i^\pm dx - \int_\Gamma \sum_i [\mu_i] J_i \, dS \\
&+ \int_\Gamma (\boldsymbol{F}_{mb} + [Q \nabla \phi]) \cdot \boldsymbol{v}_\tau dS \\
&+ \int_\Gamma (\boldsymbol{F}_{mb} \cdot \boldsymbol{n} + [F_n] - (e_\Gamma + \gamma_0) H) V_n dS \\
&+ \int_\Gamma \left[ F_\rho \right] U dS + \int_\Gamma [\phi] \frac{d^n}{dt} (C_m[\phi] - Q) \, dS + I_b
\end{aligned} \qquad (29)$$

$I_b = I_{1b} + I_{2b}$, chemical potential $\tilde{\mu}_i$ reduces to $\mu_i$; $F_n = Q\phi H$ is the same as in Eq. (17); and

$$F_\rho = \boldsymbol{n} \cdot (\boldsymbol{\sigma}_\eta + \boldsymbol{\sigma}_e) \cdot \boldsymbol{n} + \left( e_{es} + \sum_i k_B T c_i + Q \, \partial_n \phi \right). \qquad (30)$$

**Model for a deformable membrane.** To summarize, the model for fluid passing through a deformable membrane is

$$\begin{cases}
-\nabla \cdot (\varepsilon_0 \varepsilon_r^\pm \nabla \phi^\pm) = \sum_i z_i e c_i^\pm, & \text{in } \Omega^\pm \\
\frac{\partial c_i^\pm}{\partial t} + \nabla \cdot (\boldsymbol{u}^\pm c_i^\pm) = \nabla \cdot (D_i^\pm (\nabla c_i^\pm + \frac{z_i e}{k_B T} c_i^\pm \nabla \phi^\pm)), & \text{in } \Omega^\pm \\
\rho^0 \frac{d\boldsymbol{u}^\pm}{dt} + \nabla p^\pm = \nabla \cdot (2\eta^\pm \boldsymbol{D}_\eta^\pm) + \nabla \cdot (\varepsilon_0 \varepsilon_r^\pm \nabla \phi^\pm \otimes \nabla \phi^\pm - \frac{\varepsilon_0 \varepsilon_r^\pm}{2} |\nabla \phi^\pm|^2 \boldsymbol{I}), & \text{in } \Omega^\pm \\
\nabla \cdot \boldsymbol{u}^\pm = 0, & \text{in } \Omega^\pm
\end{cases} \qquad (31)$$

where the viscous stress for incompressible fluid is

$$\boldsymbol{\sigma}_\eta^\pm = 2\eta^\pm \boldsymbol{D}_\eta^\pm - p^\pm \boldsymbol{I},$$

and pressure $p$ is introduced as a Lagrange multiplier of imcompressiblity.
By taking $\bar{\rho} = \rho^0$, the interface boundary conditions on $\Gamma$ are

$$\begin{cases} \dfrac{d\boldsymbol{X}}{dt} = \boldsymbol{v} \\ -\varepsilon_0 \varepsilon_r^\pm \nabla \phi^\pm \cdot \boldsymbol{n} = C_m[\phi], \\ \boldsymbol{j}_i^\pm \cdot \boldsymbol{n} + c_i^\pm (\boldsymbol{u}^\pm - \boldsymbol{v}) \cdot \boldsymbol{n} = \dfrac{g_i}{(z_i e)^2}([k_B T(\ln(c_i/c_0) + 1) + z_i e\phi]), \\ \boldsymbol{u}^\pm - \boldsymbol{v} = U\boldsymbol{n}, \\ U = -K(x)[O + \boldsymbol{n} \cdot \boldsymbol{\sigma}_\eta \cdot \boldsymbol{n}], \\ [\boldsymbol{\sigma}_\eta + \boldsymbol{\sigma}_e] \cdot \boldsymbol{n} = (\gamma_0 - e_\Gamma)H\boldsymbol{n} - \nabla_\Gamma e_\Gamma, \end{cases} \quad (32)$$

and some appropriate boundary conditions on $\partial\Omega$ describing the experimental setup in a reasonably realistic, but idealized way. Note that the system (31) with interface condition (32) are same as the previous results[106].

**Remark** *Other formulations could be used for the membrane of course. Even the tradtional Goldman–Hodgkin–Katz flux[106]*

$$J_i = P_i f(z_i e[\phi]/k_B T) c_i^- (e^{[\mu_i]/k_B T} - 1) \text{ with } f(x) = \frac{x}{e^x - 1} \geq 0, \text{ for any } x,$$

*could be used to replace the Hodgkin-Huxley flux* $J_i = \dfrac{g_i}{(z_i e)^2}[\mu_i]$.

*What is really needed of course is a physical model of the water, mass, and electrical flux through the **ensemble** of channels, starting with atomic detail. That is not yet available, particularly because the macroscopic representation (of the ensemble of channels) needed here (i.e., in eq. (32)) should include time dependent gating phenomena that are not properties of a single permanently open channel. The Goldman-Hodgkin-Katz and Poisson Nernst Planck formulations are representations of the behavior of **single permanently open** channels. Those representations need to be supplemented by an explicit theory of time (and agonist or voltage) dependent opening and closing of **both single channels and ensembles of single channels** before they can sensibly be used in a model like eq. (32). Note two types of gating are involved, the stochastic open and closed gating of single channels, and the deterministic time dependent conductance called gating in the Hodgkin Huxley formulation of ensemble properties.*

# Permeable and Inflexible Membranes

In this section, we further assume that the membrane is inflexible, i.e. $\boldsymbol{v} = \boldsymbol{0}$. Similarly, the different between flexible and inflexible membrane is that we do not need the force balance equation on the interface. Then we could obtain the following model for fluid passing through a inflexible membrane

$$\begin{cases} -\nabla \cdot (\varepsilon_0 \varepsilon_r^{\pm} \nabla \phi^{\pm}) = \sum_i z_i e c_i^{\pm}, & in\ \Omega^{\pm} \\ \dfrac{\partial c_i^{\pm}}{\partial t} + \nabla \cdot (\boldsymbol{u}^{\pm} c_i^{\pm}) = \nabla \cdot (D_i^{\pm}(\nabla c_i^{\pm} + \dfrac{z_i e}{k_B T} c_i^{\pm} \nabla \phi^{\pm})), & in\ \Omega^{\pm} \\ \rho^0 \dfrac{d\boldsymbol{u}^{\pm}}{dt} + \nabla p^{\pm} = \nabla \cdot (2\eta D_\eta^{\pm}) + \nabla \cdot (\varepsilon_0 \varepsilon_r^{\pm} \nabla \phi^{\pm} \otimes \nabla \phi^{\pm} - \dfrac{\varepsilon_0 \varepsilon_r^{\pm}}{2} |\nabla \phi^{\pm}|^2 \boldsymbol{I}), & in\ \Omega^{\pm} \\ \nabla \cdot \boldsymbol{u}^{\pm} = 0, & in\ \Omega^{\pm} \end{cases} \quad (33)$$

where the viscous stress for incompressible fluid is

$$\boldsymbol{\sigma}_\eta^{\pm} = 2\eta^{\pm} \boldsymbol{D}_\eta^{\pm} - p^{\pm} \boldsymbol{I}$$

with interface boundary conditions on $\Gamma$

$$\begin{cases} -\varepsilon_0 \varepsilon_r^{\pm} \nabla \phi^{\pm} \cdot \boldsymbol{n} = C_m[\phi], \\ \boldsymbol{j}_i^{\pm} \cdot \boldsymbol{n} + c_i^{\pm} \boldsymbol{u}^{\pm} \cdot \boldsymbol{n} = \dfrac{g_i}{(z_i e)^2} ([k_B T(\ln(c_i/c_0) + 1) + z_i e \phi]), \\ \boldsymbol{u}^{\pm} = U\boldsymbol{n}, \\ U = -K(\boldsymbol{x})[0 + \boldsymbol{n} \cdot \boldsymbol{\sigma}_\eta \cdot \boldsymbol{n}], \end{cases} \quad (34)$$

and some appropriate boundary conditions on $\partial \Omega$.

# Historical Comments

Osmosis occurs in so many contexts that no one, certainly not us, can grasp its totality. The temptation then is to discuss only what we have grasped and say nothing about what is beyond our reach. Because we do know something about some important applications—and think something is better than nothing—we write this section. We hope workers in different fields will learn of each other's work, so they can benefit from each other's knowledge, experience, and enthusiasm about osmosis.

It is important also to understand that at least in biology many of the most important issues in osmosis remain open. We believe, without proof, that some of these issues will be easier to resolve when ionic solutions are treated as complex fluids with mathematics that is consistent, with all variables satisfying all field equations and interface boundary conditions with one set of unchanging parameters. We understand, however, that no one can know the consequences of a consistent analysis until it is actually performed in the experimental, technological and biological systems of interest. Details matter!

An enormous amount of work is being done on the theory of technological osmosis, i.e., mathematics of fluidics. Our knowledge of fluidics is too modest to allow thoughtful citation here: fortunately these applications are easy to locate with literature search methods. A search today (June 2, 2018) found two significant new papers just this week.

The classical biological literature is older, and in danger of being lost to modern generations, particularly those who know more of Navier Stokes and fluidics than of the kidney or epithelia. So we provide key references to interesting biological applications. It is important to understand that these areas are central to a wide range of biological research. They are not isolated special cases.

Much of classical physiology, described in detail in textbooks[21-24] concerns organs that depend on osmosis. Mori[106, 119] has started a consistent analysis of osmosis and our work should be viewed as an extension of his.

The reviews of Boulpaep[120], AE Hill[121], and Pohl[122] provide good entries to the field. Hill is particularly useful for showing the substantial controversies and their history. The analysis of osmosis in the lens, mostly from the laboratory of Richard Mathias, is notable for its success, combining field theory, molecular biology, structural biology, and measurements of hydrostatic pressure, electrical impedance, and fluxes to provide a coherent view of how the lens uses osmosis to stay alive.[123-125] That approach, expanded perhaps to explicitly use the theory of complex fluids, will help to resolve the many controversies we believe.

A great deal of work has been done on fusion of vesicles to membranes, because of its wide biological role (e.g., in the $\sim 10^{14}$ chemical synapses in our brains, with thousands of vesicles in each synapse) and medical importance in the entry of viruses into cells.[116, 117, 126, 127] Most of that work only considers the elastic properties of membranes. We suggest that energy sources like diffusion, electrical potential, and convection should be included in the analysis of vesicle fusion—and membrane flow in general—to see if they are used by biology for its purposes. Experiments suggest they are.

Modern work in molecular biology and water flow is focused on the aquaporins that are thought to provide channels for water flow[128, 129], or sensors controlling water flow[130, 131]. Some work seeks to study osmosis in individual channels, but comparison with similar work studying current flow through channels suggests that much remains to be done. In our view, the full range of forces—electrical, diffusional, convective, steric, and chemical—are likely to be involved in flow through water channels and so must be included in models and simulations, at least to begin with, until we discover the variables that evolution uses to control water flow on the atomic scale in the channels of aquaporins.

Sadly, osmosis and fluid flow in plants is not an area we know enough about to cite intelligently. Osmosis/fluidics reaches its most impressive heights in trees (nearly 100 meters in redwoods). Water flow in the xylem of trees—from roots to leaves in the crown—certainly needs analysis with the theory of complex fluids, if that has not already been done. There are also innumerable examples of osmosis/fluidics we do not know enough about to even name. All the more reason for readers interested in complex fluids to learn their names and study how they work.

In each case, it seems clear to us that treating osmosis as the movement of a complex fluid is likely to be useful, even necessary, in biology, and everywhere else.

# Appendix A: Proof of Lemma

Proof. By using the Batchelor's method[115, 132], we have

$$\frac{d}{dt}(\mathbf{n})dS + \frac{d}{dt}(dS)\mathbf{n} = \frac{d}{dt}(\mathbf{n}dS) = (\nabla \cdot \mathbf{v})\mathbf{n}dS - (\nabla \mathbf{v}) \cdot \mathbf{n}dS$$

$$= (\nabla \cdot \mathbf{v}\mathbf{n} - (\partial_n \mathbf{v} \cdot \mathbf{n})\mathbf{n})dS - \nabla_\Gamma \mathbf{v} \cdot \mathbf{n}dS = (\nabla_\Gamma \cdot \mathbf{v})\mathbf{n}dS - \nabla_\Gamma \mathbf{v} \cdot \mathbf{n}ds$$

and $\frac{d}{dt}dS = \nabla_\Gamma \cdot \mathbf{v}\, dS$.

Then we have $\frac{d\mathbf{n}}{dt} = -\nabla_\Gamma \mathbf{v} \cdot \mathbf{n}$.

Finally, by the definition of $\frac{d^n}{dt}$, we have

$$\frac{d^n \mathbf{n}}{dt} = \frac{d\mathbf{n}}{dt} - (\mathbf{v}_\Gamma \cdot \nabla_\Gamma)\mathbf{n} = -\nabla_\Gamma \mathbf{v} \cdot \mathbf{n} - (\mathbf{v}_\Gamma \cdot \nabla_\Gamma)\mathbf{n}$$

$$= -\nabla_\Gamma (v_n \mathbf{n} + \mathbf{v}_\Gamma) \cdot \mathbf{n} - \nabla_\Gamma \mathbf{n} \cdot \mathbf{v}_\Gamma = -\nabla_\Gamma v_n - \nabla_\Gamma (\mathbf{n} \cdot \mathbf{v}_\Gamma) = -\nabla_\Gamma v_n.$$

Here we used the symmetry of curvature tensor $\nabla_\Gamma \mathbf{n}$ and $|\mathbf{n}|^2 = 1$.

# Appendix B: Membrane mechanical properties

Membrane mechanical properties can be included in the analysis by modify Eq. (7). Assuming that $E_m$ is the energy induced by mechanical properties of the membrane, the fourth term of Eq. (7) $I_4$ yields

$$I_4 = \frac{d}{dt}\int_{\Gamma(t)}(\gamma_0 + C_m[\phi]^2)dS + \frac{dE_m}{dt}$$

$$= \int_\Gamma C_m[\phi]\frac{d^n}{dt}([\phi])\,dS - \int_\Gamma ((e_\Gamma H\mathbf{n} - \mathbf{F}_m))\cdot \mathbf{v}dS. \quad (B1)$$

where we used the results [87] $\frac{dE_m}{dt} = \int_{\Gamma(t)}\frac{\delta E_m}{\delta \Gamma}\cdot \mathbf{v}dS$ and denotes $\mathbf{F}_m = \frac{\delta E_m}{\delta \Gamma}$

Then the rate of total energy change is

$$\frac{dE^{tot}}{dt} = -\sum_i \int_{\Omega^\pm}\boldsymbol{\sigma}_\eta^\pm : \nabla \mathbf{u}^\pm dx$$

$$-\sum_\pm \int_{\Omega^\pm}\left(\boldsymbol{\sigma}_e^\pm - \epsilon_0\epsilon_r^\pm\left(\nabla\phi^\pm \otimes \nabla\phi^\pm - \frac{|\nabla\phi^\pm|^2}{2}\mathbf{I}\right)\right):\nabla \mathbf{u}^\pm dx$$

$$+\sum_\pm \int_{\Omega^\pm}\sum_i \nabla\tilde{\mu}_i^\pm \cdot \mathbf{j}_i^\pm dx - \int_\Gamma \sum_i [\tilde{\mu}_i]J_i\,dS$$

$$+\sum_\pm \int_{\Omega^\pm}\left(-\sum_i K_B T c_i^\pm + p_c^\pm\left(1 - \sum_i \frac{\partial_i\rho^\pm}{\rho^\pm}c_i\right)\right)\nabla\cdot \mathbf{u}^\pm dx$$

$$+\int_\Gamma (\mathbf{F}_{mb} - \mathbf{F}_m + [Q\nabla\phi])\cdot \mathbf{v}_\tau dS$$

$$+\int_\Gamma (\mathbf{F}_{mb}\cdot\mathbf{n} + [F_n] - (e_\Gamma H - \mathbf{F}_m\cdot\mathbf{n}))v_n dS$$

$$+\int_\Gamma \left[\frac{F_\rho}{\rho}\right]Q_\rho dS + \int_\Gamma [\phi]\frac{d^n}{dt}(C_m[\phi] - Q)\,dS - I_b \quad (B2)$$

Then we obtain the force balance equation on a membrane with mechanical properties.

$$[\boldsymbol{\sigma}_\eta + \boldsymbol{\sigma}_e]\cdot\mathbf{n} - [Q_\rho \mathbf{u}] = -e_\Gamma H\mathbf{n} - \nabla_\Gamma e_\Gamma + \mathbf{F}_m. \quad (B3)$$